\documentclass[useAMS,usenatbib]{mn2e}
\usepackage{graphicx}
\usepackage{amsmath}
\usepackage{tabularx}

\usepackage{times}
\usepackage{amssymb,amsmath}

\def\lsim{ \lower .75ex\hbox{$\sim$} \llap{\raise .27ex \hbox{$<$}} }
\def\gsim{ \lower .75ex \hbox{$\sim$} \llap{\raise .27ex \hbox{$>$}} }

\newcommand{\bi}{\begin{itemize}}
\newcommand{\ei}{\end{itemize}}

\title[Shear acceleration model for X-ray jets] 
{Constraining the shear acceleration model for the X-ray emission of large-scale extragalactic jets} 

\author[Tavecchio F.]
{F. Tavecchio$^1$\thanks{E--mail: fabrizio.tavecchio@inaf.it}
\\
$^1$INAF -- Osservatorio Astronomico di Brera, via E. Bianchi 46, I--23807 Merate, Italy\\
}

\begin{document}



\maketitle

\begin{abstract} 
The nature of the intense X-ray emission from powerful extragalactic  jets at large ($>1$ kpc) scale is still debated. The scenario that invokes the inverse Compton scattering of the CMB by electrons is challenged by the lack of gamma-ray emission in the GeV band. An alternative assumes synchrotron emission by a distinct population of ultra-high energy electrons. Here we present a concrete attempt to apply this scenario, exploring the specific model in which the ultra-high energy electrons are accelerated in a shear layer surrounding the jet. We limit the study to non-relativistic flows and particle acceleration is treated by a Fokker-Planck equation. The observed relation between low energy (radio, optical) and X-ray emission prompts us to assume that the required population of pre-accelerated particles is provided by a shock responsible for the acceleration of the electrons emitting at low frequencies. We apply the model to the emission of the principal knots of the jets of PKS 0637-752 and PKS 1136-135, two of the best studied objects. For the set of fiducial parameters adopted, the condition that the jet power does not exceeds a limiting value of $10^{48}$ erg s$^{-1}$ constrains the magnetic field above $10$ $\mu$G and indicates moderate beaming ($\delta\simeq 2$) for PKS 0637-752. For both sources, the requirement that  acceleration of the electrons proceeds faster than radiative cooling can be met if the magnetic turbulence in the shear layer follows a Kolmogorov spectrum, $I(k)\propto k^{-q}$ with $q=5/3$, but cannot satisfied in the Bohm-like case ($q=1$). 
\end{abstract}


\section{Introduction}

Extragalactic relativistic jets -- expelled as collimated outflows from the core of active galaxies and propagating for hundreds of kpc -- represent ideal laboratories to investigate the complex physics of relativistic plasmas and magnetic fields (e.g. Blandford et al. 2019). The huge difference in spatial scales (up to a factor $10^8$) and physical conditions along the jet make difficult to obtain a global view of the underlying processes and dynamics. Inner region of jets (distances from the central black hole smaller than few pc) are best studied in blazars, sources for which, due to the favorable geometry, the emitted radiation is strongly amplified by relativistic effects (e.g. Romero et al. 2017). At the opposite side of spatial scales, multiwavelength observations of extended jets provide essential information on the global dynamics of jets, their interaction with the external environment and the impact on large-scale structures (e.g. Fabian 2012).

Among the several open issues connected to jets, the nature of the localized and intense X-ray emission from knots at large ($>1$ kpc) scale in jets of powerful quasars (e.g. Sambruna \& Harris 2012) received great attention in the recent past. X-ray emission from few jets associated to bright radio-loud quasars was discovered with {\it Einstein} and {\it Rosat} observations (e.g. 3C73, Harris \& Stern 1987; R{\"o}ser et al. 2000). However, one had to wait for the superb angular resolution of {\it Chandra} to perform dedicated surveys and study the morphology and the connection with emission at radio and optical frequencies (Sambruna et al. 2002, 2004; Marshall et al. 2005). 

If interpreted as thermal or synchrotron self-Compton (SSC) radiation from non-relativistic (or mildly relativistic) jets, the observed emission requires quite unlikely and extreme physical parameters (e.g. Schwartz et al. 2000). Tavecchio et al. (2000) and Celotti et al. (2001) suggested that the observed X-ray emission could be produced via the inverse Compton (IC) scattering of the cosmic microwave background (CMB) radiation field. In order to work, this IC/CMB model requires that powerful jets exhibiting luminous X-ray emission are still highly relativistic (bulk Lorentz factor of the order of 10) and very well aligned toward the observer, so that the detected emission is characterized by a Doppler beaming factor of the order of $\delta=10$ (Tavecchio et al. 2000, Atoyan \& Dermer 2004). In this scheme the X-ray emission would be produced by slowly cooling electrons with small Lorentz factors ($\gamma\sim 10$), requiring some extra-cooling mechanism to produce the observed localized emission (Tavecchio et al. 2003, Lucchini et al. 2017). Proposed to explain the puzzling emission from the jet of the quasar PKS 0637-752 (Chartas et al. 2000, Schwartz et al. 2000), the model has been subsequently applied to interpret the emission from several jets discovered serendipitously (e.g. Simionescu et al. 2016) or in dedicated surveys (e.g. Sambruna et al. 2002, 2004, Marshall et al. 2005, 2011, 2018). In this model the physical parameters describing the emitting region can be relatively well constrained by the observational data (especially if one assumes equipartition). This allows one to draw robust predictions. In particular, one of the most important consequences of this scheme concerns the frequency and the luminosity of the peak of the IC/CMB component in the spectral energy distribution of the knots. Indeed, the peak is expected to occur in the $\gamma$-ray band, with fluxes accessible to {\it Fermi}-LAT, thus providing a powerful test-bed for the model (Tavecchio et al. 2006, Meyer \& Georganopoulos 2014). This approach has been used on the brightest sources, for which the predicted flux would be easily revealed during the quiescent phases of the bright cores, but only constraining upper limits have been derived (Meyer et al. 2014, 2015, 2017, Breiding et al. 2017).

The challenges posed to the IC/CMB scenario have stimulated the search for alternative models. A direct possibility is that the high energy component belongs to the synchrotron emission of a ultra high-energy population of electrons distinct from the population responsible for the radio-optical emission (e.g. Harris \& Krawczynski 2002, Atoyan \& Dermer 2004, Kataoka \& Stawarz 2005). This scenario is supported by observations of jets with complex morphology (e.g. Jester et al. 2002, Siemiginowska et al. 2007) and could explain the large optical polarization measured in few cases in which the optical emission appears to belong to the high-energy component (Cara et al. 2013, Perlman et al. 2020). With magnetic fields of the order of tens of $\mu$G generally assumed for jets at these distances from the core, emission in the X-ray band requires electrons with extreme Lorentz factors of the order of $\gamma=10^8-10^9$, corresponding to energies of the order of 100 TeV (e.g. Mondal \& Gupta 2019). A possibility, already explored by Stawarz \& Ostrowski (2002), is that the high-energy component is produced in a highly turbulent layer surrounding the outflow, where stochastic acceleration by turbulence could in principle push the particles to the required energies. In a similar approach (Rieger 2019 and references therein) one assumes that the acceleration proceeds through the scattering of particles in a shear layer characterized by a strong radial gradient of the velocity. In these conditions, particles diffusing in the layer experience a continuous energy gain through the scattering by the turbulence moving with different speeds. Since in this scheme the acceleration rate, proportional to the velocity difference of the scattering centers, increases for increasing mean free path of the particles, the efficiency is larger for the particles at the highest energies, while quite long acceleration times are required for low-energy particles. For this reason, this mechanism can work only if particles are pre-accelerated at sufficient energy. Liu, Rieger \& Aharonian (2017) suggest that these particles can be heated by turbulence. PIC simulations by Sironi et al. (2020) show that particle can be pre-accelerated by magnetic reconnection triggered by Kelvin-Helmholtz instability at the jet/environment interface. 
 
In this work we would like to report a first step toward the concrete application of the shear acceleration scheme to the modeling of real data. As mentioned above, one of the striking feature of the X-ray emission observed in the majority of jets is the the strict relation (although not necessarily precise spatial coincidence, see e.g. Kataoka et al. 2008, Clautice et al. 2016) with the emission at low energy. The morphological and spectral properties of the radio-optical emission strongly suggest that it is associated to shocks in the jet (e.g. Sambruna \& Harris 2012).  In the framework of dual-population models described above one has therefore to explain the strict connection between shocks and the mechanism responsible for the population of ultra-relativistic particles. With this motivation,  we develop a specific model for the multifrequency emission of knots based on the shear acceleration scenario, assuming that the required pre-heated electrons are those accelerated by the shock (sect. 2). In sect. 3 we apply the model to reproduce the emission of two well studied jets, trying to infer and constrain the physical quantities of the jet. In sect. 4 we discuss the results.

Throughout the paper, the following cosmological  parameters are assumed: $H_0=70{\rm\;km\;s}^{-1}{\rm\; Mpc}^{-1}$, $\Omega_{\rm M}=0.3$, $\Omega_{\Lambda}=0.7$.

\section{The model}

\subsection{General set-up}

As anticipated, we assume that the emission we observe from knots is produced by relativistic electrons energized by two distinct processes, namely shock acceleration (emission from radio up to optical bands) followed by the shear acceleration of some of the most energetic particles from the first population, leading to the production of a very-high energy electrons responsible for the X-ray emission (see also Merten et al. 2020 in a different context). As we will see, the timescale for the shear acceleration characteristically {\it decreases} with increasing particle energy, and therefore, this mechanism can be efficient only if a population of particles with enough energy already exists in the system. Liu et al. (2017) assumes that stochastic acceleration provides the seeds for the subsequent shear process. Recently, Sironi et al. (2020) showed by means of PIC simulations that particles can be pre-accelerated at reconnection layers generated in the jet boundary by Kelvin-Helmoltz instability. Another possibility is that the pre-accelerated particles are provided by a shock in the flow. To be specific, in the following we explore this latter scenario, although our treatment can be easily extended to the other possible pre-acceleration mechanisms. We remark that our framework offers a natural way to explain why the X-ray emission is always associated to low-energy radiation, not obvious in other schemes. 

\begin{figure}
 \centering
 \hspace*{-0.25truecm}
 \vspace*{0.5 truecm}
 \includegraphics[width=0.47\textwidth]{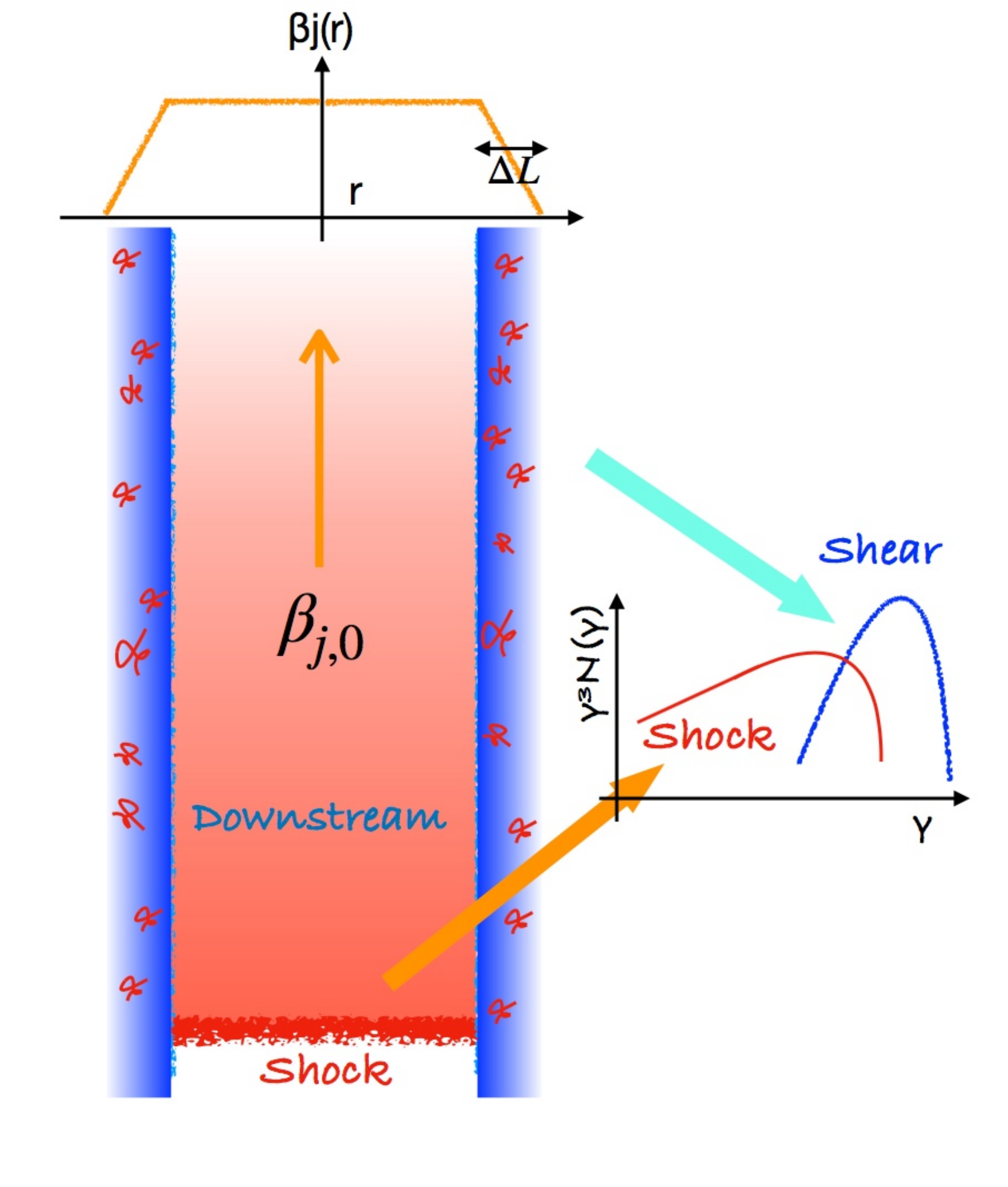}\\
\vspace*{-0.5truecm}
 \caption{Sketch of the geometry assumed (not to scale). At a shock front particles are accelerated following a power-law energy distribution and are advected in the downstream region, where they emit through synchrotron and IC mechanisms. A fraction of these relativistic electrons diffuses into the sheared flow surrounding the jet (blue region with width $\Delta L$), where further acceleration occurs through the mediation of the the turbulence (red), resulting into a bump-like distribution at larger energies, responsible for the X-ray emission.}
 \label{fig:cartoon}
\end{figure}

A sketch of the geometry is reported in Fig. \ref{fig:cartoon}.
We assume the existence of a shock, possibly due to recollimation from external gas (e.g. Komissarov 1994).  Through the standard diffusive shock acceleration process (e.g. Blandford \& Eichler 1987) particles are accelerated and injected downstream. This is the region where the electrons loose energy through synchrotron and IC emission and that observationally is identified as the emission knot. The spectral shape of the emission from knots traced by radio, IR and optical data is well described by synchrotron radiation of a population of electrons with a cut-offed power law energy distribution. Consistently, we assume that the electron population deriving from shock acceleration is described by the following energy distribution:
\begin{equation}
n_0(\gamma)=K \gamma^{-n_{\rm sh}} \exp\left(-\frac{\gamma}{\gamma_{\rm cut}} \right); \;\;\; \gamma> \gamma_{\rm min}   
\label{eq:n0}
\end{equation}
where $K$ is a normalization and $\gamma_{\rm cut}$ is the maximum Lorentz factor reached by the particles. For the magnetic field intensities inferred for these regions, of the order of 10 $\mu$G, the radiative cooling time of the particles, even close the maximum energy, is very long. This allows us to assume that the electron population coming from the shock is stationary (we assume that equilibrium is assured by continuous escape/advection from the emission region).

We assume that the jet is surrounded by a shear layer of thickness $\Delta L$ where the bulk velocity of the plasma $\beta_{j}(r)$ decreases with the distance from the jet axis, $r$. We further assume that the plasma supports turbulence against which particle can resonantly scatter. Due to the difference speed of the plasma, a particle moving in the radial direction experiences scatterings with centers of different speeds, and statistically gains energy after multiple scatterings (e.g. Rieger 2019). Under the action of this process scattering the electrons can reach Lorentz factors up to $\gamma \simeq 10^8-10^9$ and emit synchrotron radiation in the X-ray band. The distribution resulting from the action of the shear acceleration process is calculated using the treatment developed in Liu et al. (2017), as sketched in Sect 2.2.

The downstream region is modeled as a cylinder with radius $r_{\rm d}=r_j-\Delta L$ and length $l=2 r_j$, where $r_j$ is the total jet radius. After a distance $\sim l$ we expect that jet expansion results in adiabatic losses effectively quenching the emission (e.g. Tavecchio et al. 2003). The shear layer instead is a hollow cylinder with internal radius $r_{\rm d}$, external radius $r_j$ and length $l$. For both regions (downstream and shear layer) we calculate the synchrotron and IC emission using the resulting electron distributions adapting the code fully described in Maraschi \& Tavecchio (2003).
For the IC scattering we use the full Klein-Nishina cross section and we consider both the synchrotron radiation and the CMB.  Due to the small bulk Lorentz factors explored in this paper, the small amplification of the radiation energy density due to the relative motion of the downstream and the shear (e.g. Ghisellini et al. 2005) can be neglected. Finally, we make the simplifying assumption that  the magnetic field has the same value throughout the downstream shock region and in the shear layer. Therefore we neglect the possible amplification of the field in the shear layer due to turbulent dynamos (e.g. Zhang et al. 2009) or the evolution of the field in the downstream region (e.g. Tavecchio et al. 2018, Baring et al. 2017). Both processes could have an important impact on the multifrequency observational properties of the knots (see Discussion below).

\subsection{Modeling the shear acceleration}

A key parameter regulating the physics of the shear acceleration is the mean free path of particles scattered by magnetic turbulence, $\lambda$, conveniently expressed as (e.g. Liu et al. 2017):
\begin{equation}
\lambda(\gamma)=\frac{r_g}{\xi} \left(\frac{r_g}{\Lambda}\right)^{1-q},
\end{equation}
where $\gamma$ is the particle Lorentz factor, $r_g$ is the particle Larmor radius, $\xi=\delta B^2/B^2$ is the ratio between the energy density of the turbulent and the regular magnetic field, $\Lambda$ is the maximum wavelength of turbulence interacting with the particles and $q$ is the slope of the power law describing the energy spectrum of the turbulent field, $I_B(k)\propto k^{-q}$. In the following we will consider two cases, $q=1$ (corresponding to Bohm-like diffusion) and $q=5/3$ (Kolmogorov spectrum).

Having defined the jet velocity profile, $\beta_{j}(r)$, and the mean free path of particles, $\lambda(\gamma)$, one can derive the average rate of momentum change due to the scattering off the turbulent waves (we refer to Liu et al. 2017 for the complete derivation):
\begin{equation}
\langle \frac{\Delta\gamma}{\Delta t} \rangle= \frac{6-q}{15} A^2 \gamma \frac{\lambda(\gamma)}{c},
\end{equation} 
where $A$ is a parameter related to the velocity profile, $A=\Gamma_j(r)^2 \, |\partial_r \beta_{j}(r)| c$. Analogously, one can derive the average rate of momentum dispersion (related to the momentum diffusion coefficient):
\begin{equation}
\langle \frac{(\Delta\gamma)^2}{\Delta t} \rangle= \frac{2}{15} A^2 \gamma^2 \frac{\lambda(\gamma)}{c}.
\end{equation}


The acceleration timescale can be quantified with the standard relation: 
\begin{equation}
t_{\rm acc}(\gamma)=\frac{\gamma}{\langle \Delta\gamma/\Delta t\rangle}=\frac{15}{6-q}\frac{c}{\lambda A^2}.
\end{equation}
Note that $t_{\rm acc}\propto (\gamma/B)^{-1}$ for $q=1$ and $t_{\rm acc}\propto (\gamma/B)^{-1/3}$ for $q=5/3$. In both cases $t_{\rm acc}$ decreases with the particle energy, i.e. particle at large energies are more efficiently accelerated than low energy ones. This is the reason why a population of pre-accelerated electrons is required for this mechanism. On the other hand, large magnetic fields, determining small Larmor radii and thus small $\lambda$, imply large acceleration times. Note also that, for the same $\gamma$, $t_{\rm acc}$ is smaller in the Kolmogorov ($q=5/3$) case than  in the Bohm-like case, because of the larger $\lambda$.

Particles can also diffusively escape from the acceleration layer. The treatment of the escape involves the microphysics of particle transport in shear flows, matter of intense investigation (e.g. Webb et al. 2018, Rieger \& Duffy 2019). Following Rieger \& Duffy (2019) we approximated the escape time considering the diffusion of particles from a region with size comparable to the shear layer thickness $\Delta L$, i.e.:
\begin{equation}
t_{\rm esc}(\gamma)\simeq \frac{\Delta L^2}{2\kappa},
\end{equation}
where for the spatial diffusion coefficient one can use the standard expression $\kappa=c\lambda/3$. In the calculations we assume $t_{\rm esc}\simeq \Delta L^2/c\lambda$. Note that, in order to ensure acceleration, we have to assume $t_{\rm acc}<t_{\rm esc}$, implying the condition:
\begin{equation}
\frac{6-q}{15} \Gamma_j^4 \left( \frac{\partial \beta_j}{\partial r}\right)^2 \Delta L^2 \gtrsim 1.
\label{eq:tacctcool}
\end{equation}

With these quantities at hand, one can write the Fokker-Planck type equation ruling the evolution in time of the electron energy distribution $n(\gamma,t)$:
\begin{equation}
\begin{split}
\frac{\partial n(\gamma,t)}{\partial t} = \frac{1}{2} \frac{\partial}{\partial \gamma} \left[  \langle \frac{\Delta\gamma^2}{\Delta t} \rangle \frac{\partial n(\gamma,t)}{\partial \gamma} \right] \\
- \frac{\partial}{\partial \gamma} \left[\left(  \langle \frac{\Delta\gamma}{\Delta t} \rangle -\frac{1}{2}\frac{\partial}{\partial\gamma} \langle \frac{\Delta\gamma^2}{\Delta t} \rangle   + \dot{\gamma}_c\right) n(\gamma,t)\right] \\
- \frac{n(\gamma,t)}{t_{\rm esc}(\gamma)} + Q(\gamma,t),
\end{split}
\label{fp}
\end{equation}
where the term $Q(\gamma,t)$ accounts for the injection of new particles in the system and $\dot{\gamma}_c$ includes energy losses.  The relevant losses here are those due to synchrotron and IC emission, with rate:
\begin{equation}
\dot{\gamma}_c=- \frac{4}{3}\frac{\sigma_T}{m_ec} (U_B+U_{\rm rad})\beta^2\gamma^2,
\end{equation}
and cooling time:
\begin{equation}
t_{\rm cool}(\gamma)=\frac{\gamma}{ \lvert\dot{\gamma}_c \rvert}= \frac{3 m_ec}{4 \sigma_T (U_B+U_{\rm rad})\beta^2\gamma}.
\end{equation}
The total radiation energy density includes the contribution of low-energy synchrotron photons emitted by electrons in the downstream region and the CMB photons, $U_{\rm rad}=U_{\rm syn}+U_{\rm CMB}$. We include the effects of the Klein-Nishina cross section using the step-function approximation (e.g. Coppi \& Blandford 1990), defining an energy-dependent effective radiation energy density (Tavecchio et al. 1998):
\begin{equation}
U_{\rm rad,eff}(\gamma)= \int _0^{3mc^2/4h\gamma} U_{\rm rad}(\nu)d\nu.
\end{equation}

Eq. \ref{fp} can be numerically solved using standard schemes to derive the evolution of the energy distribution with time. We adopt the robust implicit method of Chang \& Cooper (1970), especially suited for Fokker-Planck type equations.

We assume that a fraction of the particles downstream of the shock, characterized by the distribution $n_0(\gamma)$ defined by Eq. \ref{eq:n0}, enters the shear acceleration process and are further accelerated to higher energies. The rate of injected particles can be phenomenologically described by an injection time $\tau_{\rm inj}$, related to the diffusion time in the downstream region. Since the details of the particle transport are far from clear, for simplicity we assume an energy independent timescale.  Therefore we assume a constant injection with spectrum $Q(\gamma)=n_0(\gamma)/\tau_{\rm inj}$. 

We have now to specify the velocity profile of the jet. A quite plausible reason for the postulated velocity stratification is the interaction of the jet with the external medium, possibly mediated by the Kelvin-Helmoltz instability (e.g. Sironi et al. 2020, Borse et al. 2020). In this condition one can assume that the jet has an unperturbed central core with constant velocity, surrounded by a sheath with a steep velocity gradient. We use the following velocity profile:
\begin{equation}
\beta_j(r)=\left\{ \begin{array}{ll}
                    \beta_{j,0}  &  \mbox{$r \leq r_j-\Delta L$} \\
		    \beta_{j,0} - \frac{\beta_{j,0}}{\Delta L} \left(r-r_j+\Delta L\right)  &  \mbox{$r> r_j-\Delta L$}
		   \end{array} 
		   \right. 
\end{equation}
where $\Delta L$ characterizes the thickness of the layer (see Fig. \ref{fig:cartoon}).
 
At very high-energy the mean free path of the electrons becomes comparable to the size of the jet layer, $\Delta L$. Beyond this point the particles escape from the system and cannot be further accelerated. In the numerical scheme this is modeled setting to zero the coefficients $\Delta\gamma/\Delta t$ and $(\Delta\gamma)^2/\Delta t$ for these energies.

\begin{figure}
 \centering
 \hspace*{-0.9truecm}
 \includegraphics[width=0.48\textwidth]{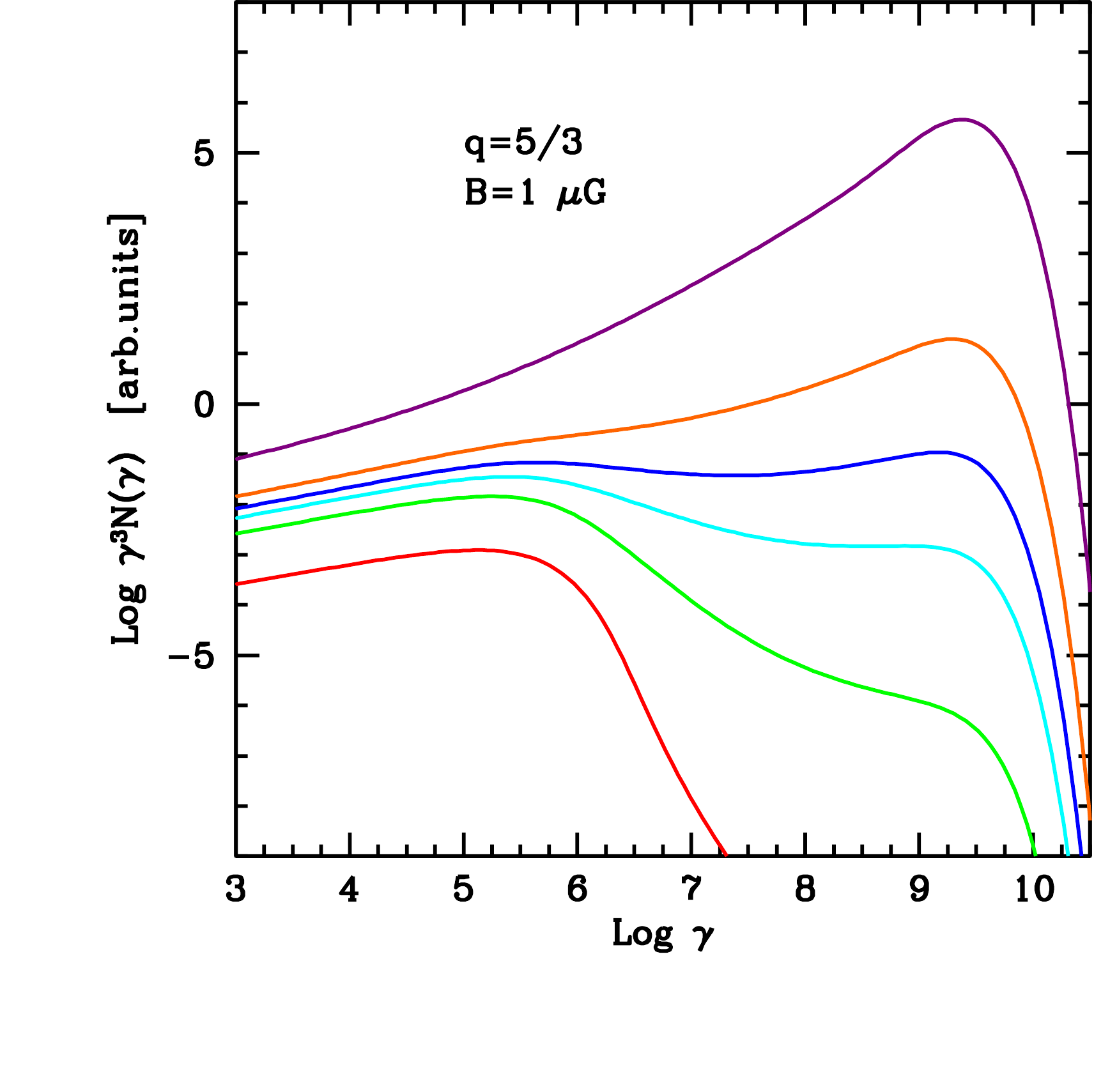}\\
 \vspace*{-.5truecm}
 \caption{Example of the evolution in time of the electron energy distribution produced by the shear acceleration. We assume a continuous injection following a cut-offed power law distribution, $Q(\gamma)\propto \gamma ^{-n_{\rm inj}}\exp(-\gamma/\gamma_{\rm cut})$ with $n_{\rm inj}=2.6$ and $\gamma_{\rm cut}=3\times 10^5$ and the curves report the resulting distribution after  (from bottom to top)  $\Delta t=5$, 50, 100,  150, 250 and $10^{3}$ (with time expressed in units of $r_j/c$). We assume $r_j=10^{21}$ cm, $B=1$ $\mu$G and $q=5/3$.}
 \label{fig:evo}
\end{figure}

As an example, in Fig. \ref{fig:evo} we show the solutions of Eq. \ref{fp} assuming $r_j=10^{21}$ cm, $B=1$ $\mu$G, $q=5/3$, $\Delta L=r_j/10$, $\xi=0.1$. The (continuous) injection follows a cut-offed power law in energy, $Q(\gamma)\propto \gamma ^{-n_{\rm inj}}\exp(-\gamma/\gamma_{\rm cut})$ with $n_{\rm inj}=2.6$ and $\gamma_{\rm cut}=3\times 10^5$. The curves report the energy distribution of the electrons at successive times (from bottom to top, in units of $r_j/c$: 5, 50, 100,  150, 250, 10$^{3}$). The curves clearly show that the acceleration process leads to the formation of a narrow peak whose maximum  at $\gamma\sim 10^9$ marks the energy where  acceleration and cooling times are equal.

\section{Application}

In this section we apply the model described above to the multifrequency emission from knots of jets associated to the benchmark sources PKS 0637-752 (the first detected by {\it Chandra}) and PKS 1136-135 (the target of several deep multifrequency observations).
\\

\begin{figure}
 \centering
 \hspace*{-0.9truecm}
 \includegraphics[width=0.45\textwidth]{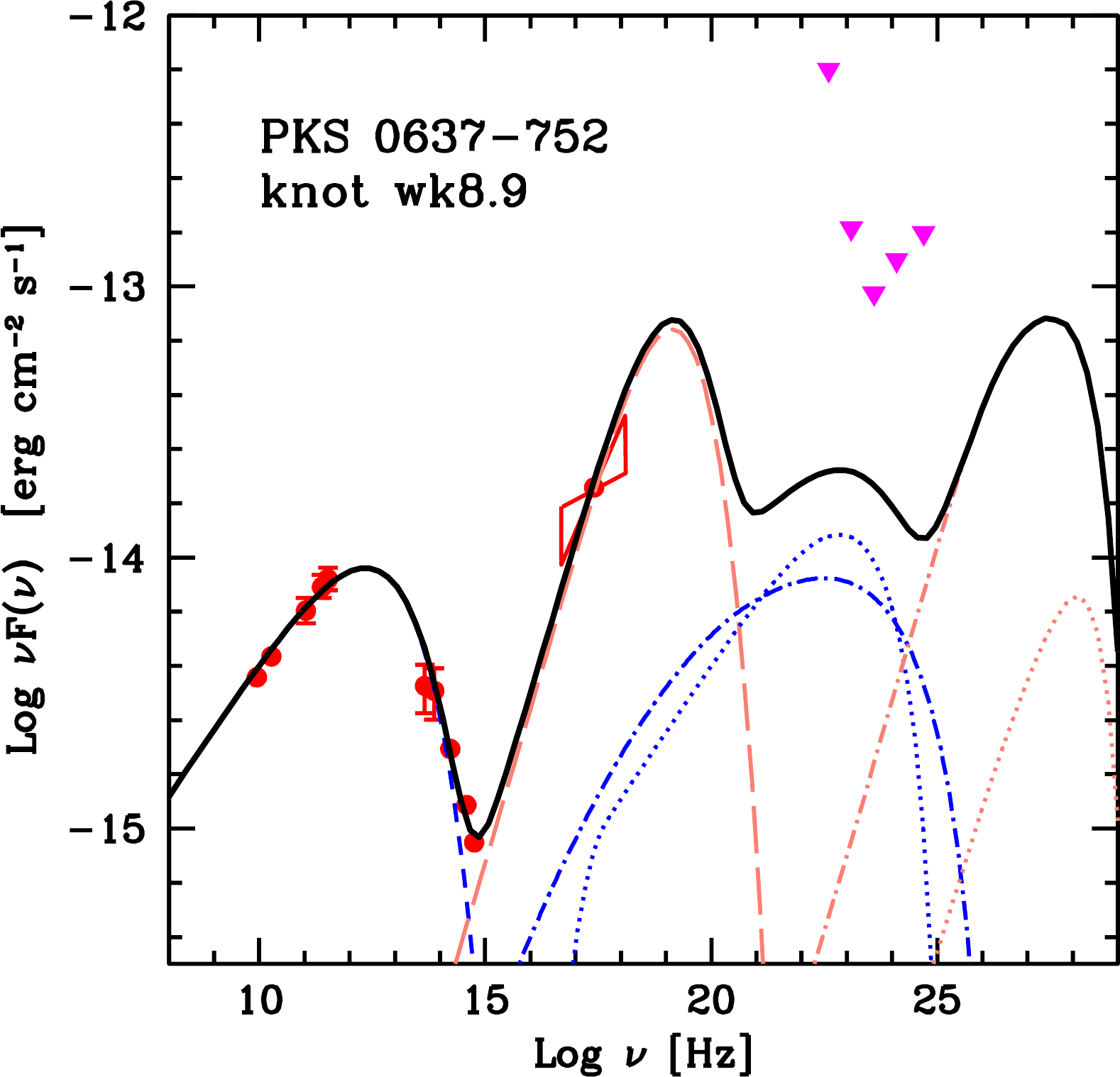}\\
 \vspace*{0.35 truecm}
 \caption{Spectral energy distribution of knot wk8.9 of PKS 0637--752. Data from Lovell et al. (2000) (ATCA), Meyer et al. (2017) (ALMA), Uchiyama et al. (2005) ({\it Spitzer}), Mehta et al. 2009 ({\it HST}), Chartas et al. 2000 ({\it Chandra}). Magenta triangles show the {\it Fermi}/LAT upper limits. We report the result of our model for $q=5/3$. The blue lines show the components emitted by the downstream shock region, orange lines are for the emission of the shear layer. Dashed: synchrotron; dotted: IC/CMB; dot-dashed: IC from synchrotron of the shock.}
 \label{fig:0637}
\end{figure}

{\it PKS 0637-752}: we use the multifrequency data of the well-defined knot wk8.9 reported by Mehta et al. (2009) and Meyer et al. (2017), see Fig. \ref{fig:0637}. This dataset is particularly reach, including, besides data at radio and X-ray frequencies, a good coverage of the low-energy emission component in the mm (ALMA), IR ({\it Spitzer}) and optical ({\it HST}) bands. These data greatly constrain the spectrum of the low energy component, well reproduced by a power law with a cut-off in the IR-optical region. 

In the application of the model several parameters can be kept fixed. In particular, the size knot can be derived from observations, $r_j \simeq 2\times 10^{21}$ cm (e.g. Tavecchio et al. 2000). We assume the fiducial values $\Delta L=0.15 r_j$, $\Lambda =\Delta L$ and  $\xi=0.1$. Since $|\partial _r\beta|\simeq \beta_{j,0}/\Delta L$, the condition (\ref{eq:tacctcool}) implies a minimum Lorentz factor $\Gamma_{\rm j,0}\gtrsim 1.6$. We adopt $\Gamma_{j,0}=1.7$.

\begin{figure}
 \centering
 \hspace*{-0.9truecm}
 \includegraphics[width=0.45\textwidth]{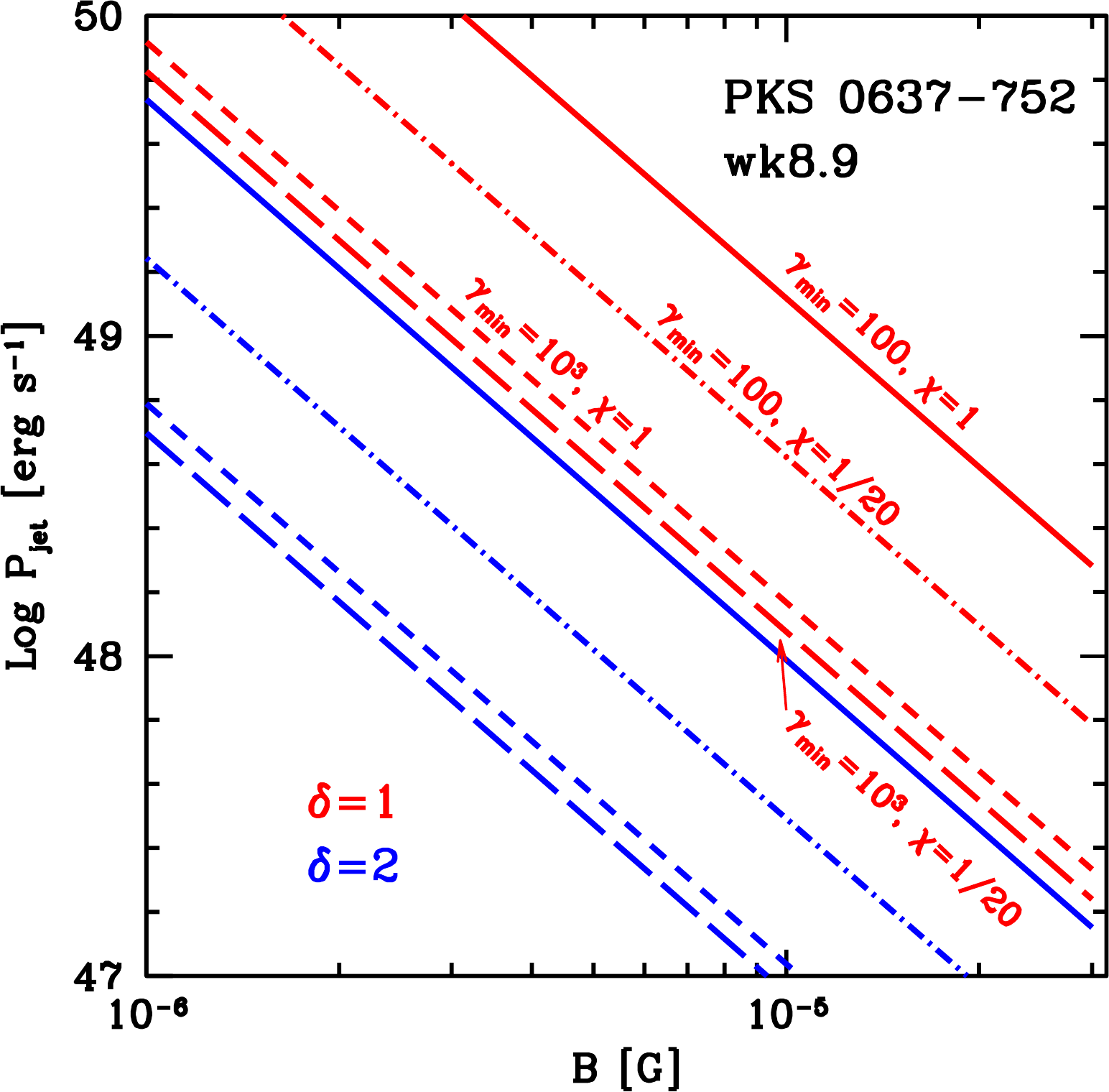}\\
 \vspace*{0.35truecm}
 \caption{Jet power as a function of the magnetic field required by the observed low-energy synchrotron emission of knot wk8.9 of PKS 0637--752. We report two sets of lines (for $\delta=1$ and 2) for normal ($\chi=1$) and pair-rich ($\chi=1/20$) plasma and for minimum Lorentz factor of the emitting electrons $\gamma_{\min}=100$ and $10^3$.}
 \label{fig:power}
\end{figure}

First of all consider the low energy component. The slope of the underlying electron population can be fixed by using the radio data. The other parameters, in particular the magnetic field and the electron normalization, cannot be determined uniquely. However, a strong constraint on the magnetic field can be derived by considering the jet energy flux (or power) required to account for the observed luminosity (see also Ghisellini \& Celotti 2001). 
Considering the outflow composed by relativistic electrons and (cold) protons (the magnetic field contribution is negligible), the energy flux can be written as (e.g. Bicknell 1994):
\begin{equation}
P_{\rm jet}=\pi r_j^2\Gamma_j^2 \beta c n_e [\langle \gamma \rangle m_e c^2 +\chi m_p c^2(\Gamma_j-1)/\Gamma_j]
\label{eq:pjet}
\end{equation}
where $n_e$ is the electron density $\langle \gamma \rangle$ the average electron Lorentz factor $\langle \gamma \rangle = \int \gamma n_e(\gamma) d\gamma/n_e$ and $\chi\equiv n_p/n_e$ (where $n_p$ is the proton density) depends on the plasma composition ($\chi=1$ for an equal number of protons and electrons). 
The specific luminosity emitted through synchrotron by electrons following a power law energy distribution (Eq. \ref{eq:n0}) can be written as (e.g. Ghisellini et al. 1985):
\begin{equation}
L(\nu)=c(\alpha) K B^{1+\alpha}\nu^{-\alpha} V \delta^{3+\alpha}
\label{eq:ls}
\end{equation}
where $c(\alpha)$ is a constant depending on the spectral slope $\alpha$, $K$ is the power law normalization, $V$ is the volume and $\delta$ the relativistic Doppler factor.
Starting from the observed flux in the radio band, one can use Eq. \ref{eq:ls} to derive $K$ and insert it into Eq. \ref{eq:pjet}, obtaining an expression for the jet power as a function of the magnetic field. The only free parameters are $\chi$ and $\gamma_{\rm min}$, the minimum Lorentz factor of the emitting electrons (which enters in the calculation of $\langle \gamma \rangle$). 

For the case of knot wk8.9 we report the derived jet power as a function of the magnetic field in Fig. \ref{fig:power} assuming that the emission is not substantially beamed, $\delta\simeq 1$ (red lines). We show the result for two different values of $\gamma_{\rm min}=10^2$ and $10^3$ and for $\chi=1$ (normal plasma) and $\chi=1/20$ (pair enriched plasma). In all cases the derived power is quite large, exceeding $P_{\rm jet}=10^{48}$ erg s$^{-1}$ for fields lower than 10 $\mu$G. For powerful radio-loud quasars the largest inferred power are around $10^{48}$ erg s$^{-1}$ (e.g. Ghisellini et al. 2014). Fig. \ref{fig:power} shows that to not exceed this value the magnetic field must be larger than $B>10$ $\mu$G even in the most optimistic cases (pair-rich jet and large $\gamma_{\rm min}$). However, as discussed below, magnetic fields substantially larger than 10 $\mu$G are unsuitable for the shear acceleration. 

A possibility to relax the situation is to assume some degree of beaming of the observed radiation. In Fig. \ref{fig:power} we report a set of curves calculated assuming $\delta=2$ (blue lines) which, for the assumed bulk Lorentz factor, implies an observing angle $\theta_{\rm v}\simeq 30$ degrees. The effect of beaming is to reduce the number of electrons required to produce the observed luminosity, thus leading to reduce the total amount of energy flux of the jet. In this case magnetic fields of few $\mu$G are allowed for power lower than $10^{48}$ erg s$^{-1}$, at least in the case of high minimum Lorentz factors or pair-rich jets. In the following we therefore fix $B=10$ $\mu$G and $\delta=2$.

\begin{figure}
 \centering
 \hspace*{-0.9truecm}
 \includegraphics[width=0.45\textwidth]{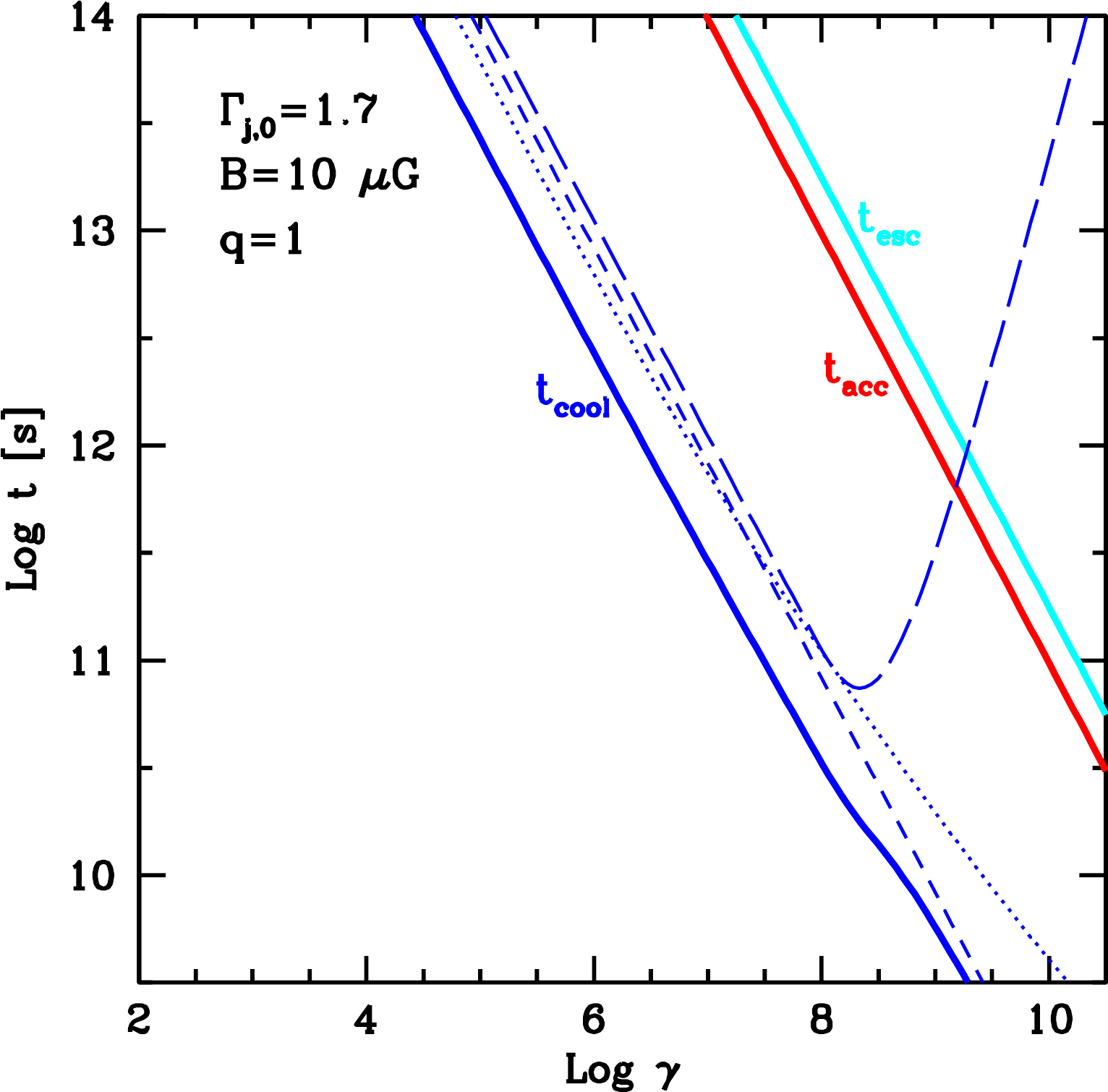}\\
 \vspace*{0.35truecm}
 \caption{Timescales relevant for the shear acceleration process as a function of the particle Lorentz factor for $q=1$. We report the acceleration (red), total cooling (blue) and escape (light blue) timescales. We also report the curves showing separately the cooling time for synchrotron, IC on the synchrotron photons and IC/CMB (dashed, dotted and long-dashed lines, respectively).}
 \label{fig:timesq1}
\end{figure}

\begin{figure}
 \centering
 \hspace*{-0.9truecm}
 \includegraphics[width=0.45\textwidth]{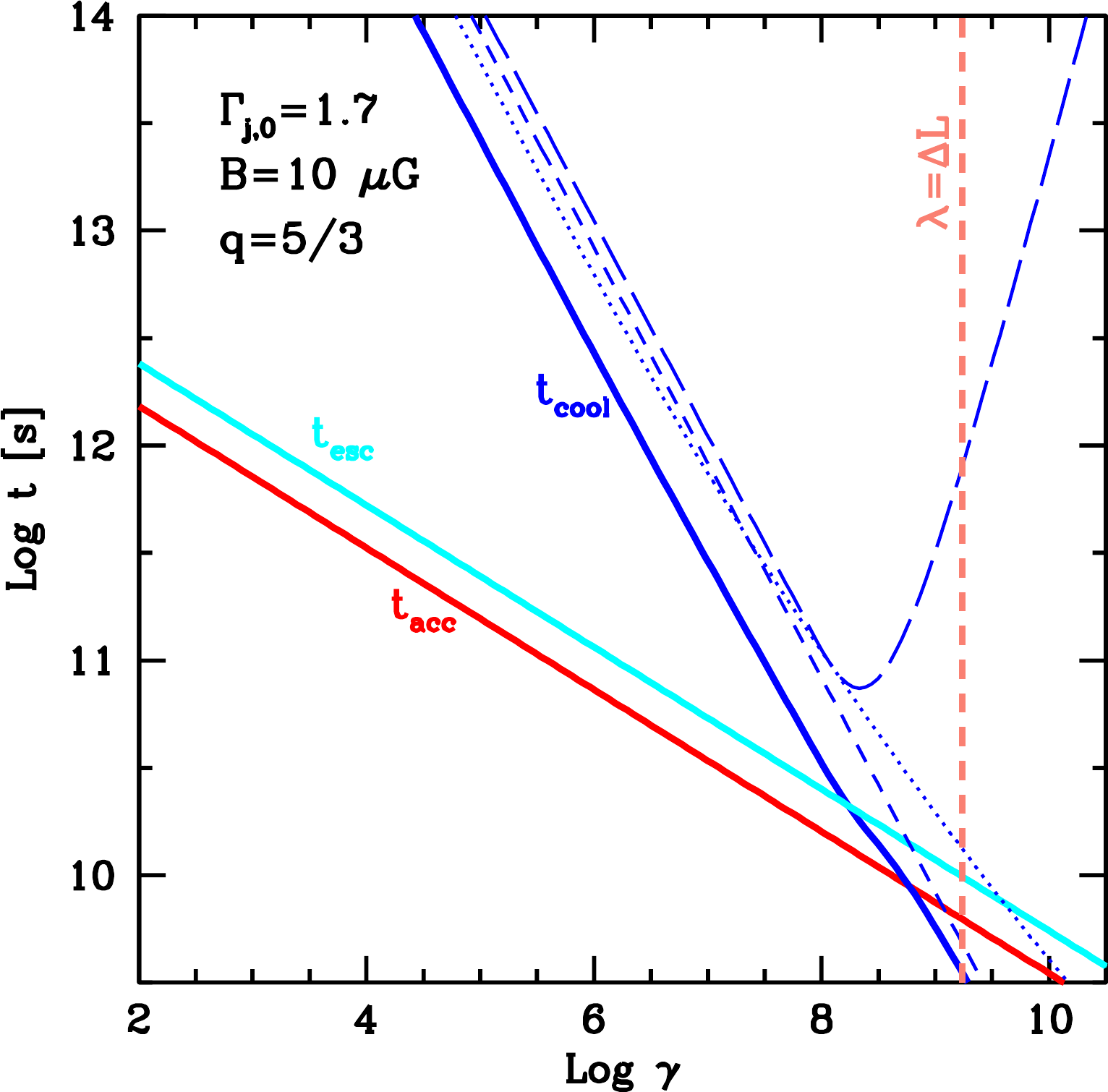}\\
 \vspace*{0.35truecm}
 \caption{As Fig.\ref{fig:timesq1} for the case $q=5/3$. The vertical orange dashed line shows the Lorentz factor above which the mean free path $\lambda$ exceeds the width of the shear layer, thus halting the acceleration process}
 \label{fig:times}
\end{figure}

Let us now consider the shear layer. In Fig. \ref{fig:timesq1} and \ref{fig:times} we show the relevant timescales involved in the shear acceleration process as a function of the particle energy assuming $B=10$ $\mu$G for $q=1$ and $q=5/3$, respectively. The solid blue line shows the total cooling time, whereas the dashed, dotted and long-dashed thin curves separately report the cooling time for synchrotron, IC on the synchrotron photons and IC/CMB. The acceleration and the escape timescales are shown by the red and light blue curves.

For $q=1$ (Fig. \ref{fig:timesq1}) at all energies the radiative cooling is clearly much faster than the acceleration, i.e. particles cannot gain energy and acceleration is ineffective. Note that IC losses are as relevant as the synchrotron ones. Therefore, a substantially reduction of the magnetic field would not have a strong impact on the total cooling time, since the radiation field are constant. On the other hand, a lower $B$ would determine a large $\lambda$ and, in turn, a smaller acceleration time. However, an effective acceleration (i.e. $t_{\rm cool}>t_{\rm acc}$) could only be achieved for $B$ well below 1 $\mu$G, implying implausible jet power exceeding $10^{49-50}$ erg s$^{-1}$.

In Fig. \ref{fig:times} we report the same set of curves for the case $q=5/3$ (i.e. Kolmogorov turbulence). In this case the reduced mean free path of the electrons ensures an acceleration much faster than cooling up to $\gamma\sim 10^9$, where cooling is dominated by synchrotron (dashed blue line). In this case it is therefore possible to accelerate electrons to the required energies, keeping the jet power to acceptable values. The acceleration timescale of the particles at the highest energy is of the order of $10^{10}$ s, i.e slightly smaller than $r_j/c$. Note that larger magnetic fields would increase $t_{\rm acc}$ and simultaneously decrease $t_{\rm cool}$, leading to a lower maximum energy of accelerated electrons. With a lower $B$, instead, electrons could potentially achieve larger Lorentz factors. However, in this case the acceleration would be limited by the confinement of the particles, expressed by the condition $\lambda < \Delta L$ (vertical orange dashed line in Fig. \ref{fig:times}). Therefore, in practice, one does not expect Lorentz factors larger than $\approx 10^9$.

In Fig. \ref{fig:0637} we present the result of the model with $\delta=2$, $B=10$ $\mu$G and $q=5/3$ (the full list of parameters is provided in Table \ref {tableparam}). The low-energy component is reproduced with  synchrotron emission by electrons following the distribution Eq. \ref{eq:n0}, matching the data to the model by varying the normalization $K$, the slope $n_{\rm sh}$ and the cut-off Lorentz factor. For definiteness we fix $\gamma_{\rm min}=300$. Note that this parameter has no influence on the observed emission (the emission of electrons at $\gamma_{\rm min}$ is concentrated below the MHz band) but has an important impact on the  the derived power (since, for a steep distribution, the total number of particles is dominated by those at the lowest energies). The IC component of the emission (dashed and dotted blue lines) peaks in the GeV band but the flux is safely below the LAT upper limits. The contribution to the total X-ray flux is also negligible.
  
\begin{table*}
\centering
\begin{tabular}{cccccccccc}
\hline
\hline
Knot  & $q$ & $\gamma_{\rm cut}$ $(\times 10^5)$ & $K$ & $n_{\rm sh}$ & $B$ & $\delta$ & $\tau_{\rm inj}$ & $t$ & $P_{\rm jet}$\\
\quad  & [1] & [2]  & [3] & [4] & [5] & [6] & [7] & [8]& [9]  \\
\hline
PKS 0637-752 wk8.9 & 5/3 & 5.3 & 1.75 & 2.5 & 10 &2 & $10^3$ & 10 & 2.6\\
PKS 1136-135 A & 5/3 & 15 & 0.2 & 2.3 & 10 &1 & $2.2\times 10^3$ & $3$ & 1.2\\
\hline
\hline
\end{tabular}
\vskip 0.4 true cm
\caption{
Parameters of the models.
[1]: power law index of the turbulent spectrum;
[2] cut-off electron Lorentz factor of the shock component;
[3]: normalization of the the shock electron energy distribution (particle cm$^{-3}$);
[4]: slope of the the shock electron energy distribution;
[5]: magnetic field ($\mu$G)
[6]: Doppler factor
[7]:  injection timescales for the shear acceleration in units of the light-crossing time $r_j/c$;
[8]: time in units of the light-crossing time $r_j/c$.
[9]: jet power ($10^{47}$ erg s$^{-1}$).}
\label{tableparam}
\end{table*}

The second emission component, produced by the electrons accelerated within the velocity shear, is matched to the X-ray data by regulating the only remaining free parameters, namely the lifetime, $t$, and the injection timescale, $\tau_{\rm inj}$. As shown in Fig.\ref{fig:evo}, a pronounced bump (peaking at the maximum electron energy determined by the balance between cooling and acceleration) develops only at sufficiently late times, when the evolution converges to a steady state distribution. Indeed, we check that, to properly reproduce the quite hard optical-X ray continuum traced by the data, the lifetime of the source has to be large enough to reach an equilibrium state. For smaller $t$, in fact, the incompletely developed bump displays a relatively soft spectrum incompatible with the optical and X-ray fluxes. Once $t$ is fixed, the spectrum can be properly normalized acting on $\tau_{\rm inj}$. 


The IC components of the shear layer emission are expected in the very-high energy band, in the energy ranged probed by Cherenkov arrays. However, the low predicted flux (even taking into account the integrated emission from the entire jet) and the severe absorption of the photons interacting with the extragalactic background light (not included here) make very difficult the possible detection by current and future instruments.
\\

\begin{figure}
 \centering
 \hspace*{-0.9truecm}
 \includegraphics[width=0.45\textwidth]{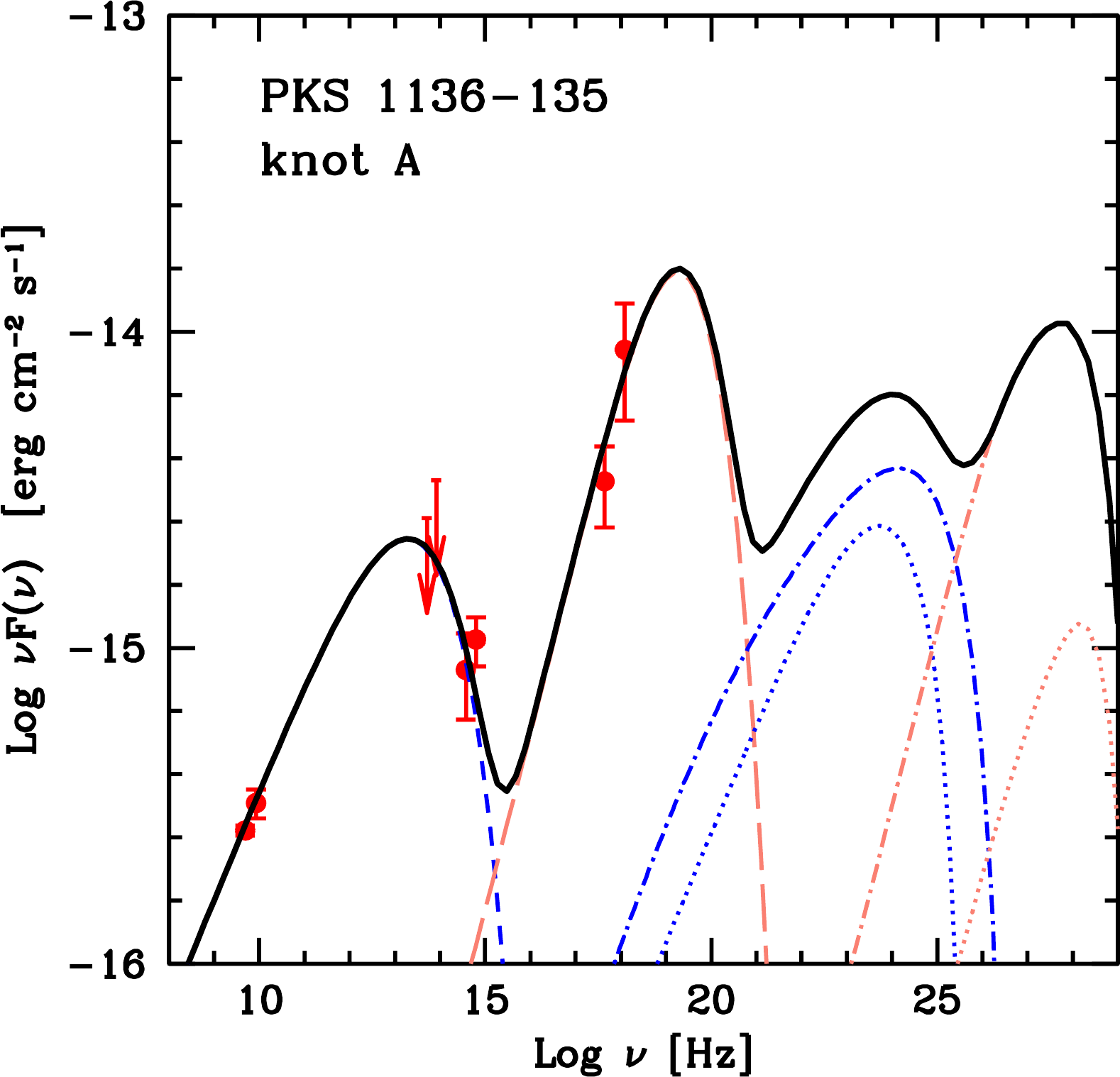}\\
 \vspace*{0.35truecm}
 \caption{Spectral energy distribution of knot A of PKS 1135-135 (data from Sambruna et al. 2002 and Uchiyama et al. 2007). Lines as in Fig.\ref{fig:0637}.}
 \label{fig:1136a}
\end{figure}

{\it PKS 1136-135}: this source has been the target of several dedicated observations in different bands. In particular, we consider the brightest feature, called knot A. Deep {\it Chandra} and {\it HST} observations (Sambruna et al. 2004, 2006) and IR data from {\it Spitzer} (Uchiyama et al. 2006, 2007) allows us to have a good description of its multifrequency emission (Fig. \ref{fig:1136a}).

The knot radius derived by Sambruna et al. (2020) is $r_j=2\times 10^{21}$ cm. As above we further fix $\Gamma_{j,0}=1.7$, $\Lambda = \Delta L$ cm, $\xi=0.1$. 

Also in this case we considered the two cases $q=1$ and $q=5/3$.
The situation is very similar to the previous case, since for $q=1$ the acceleration is too slow to counterbalance cooling (for simplicity we do not report the plots with the timescales, very similar to those discussed before). Only the case $q=5/3$ offers a viable solution. 
The jet power poses a less tight constraint than for PKS 0637-052. Indeed, even for $\delta=1$ the power can be kept to acceptable values.
The corresponding model is reported in Fig. \ref{fig:1136a} and the parameters in Table 1.
\\



\section{Discussion}

The problem posed by the origin of the intense X-ray emission observed in correspondence with knots in jets of powerful quasars is still open, more than 20 years after the first detections. The IC/CMB model is currently challenged by observational evidence, especially the lack of the expected $\gamma$-ray emission in the GeV band. Since the prediction of high-energy emission is a quite robust feature of the model (e.g. Tavecchio 2006, Georganopoulos et al. 2016, Meyer 2014), the non-detection is rather hard to explain with minor adjustments (e.g. Lucchini et al. 2017). On the other hand, it has to be remarked that, even if not dominant in misaligned low-$z$ sources, when the jet is strongly aligned with the line of sight (as assumed for blazars, e.g. Tavecchio et al. 2007, Meyer et al. 2019) or for sources located at high redshift (Simionescu et al. 2016, Worrall et al. 2020, Schwartz et al. 2020) -- where the increased CMB energy density naturally boosts the IC output (Schwartz 2002) -- the IC/CMB framework still offers a viable solution. 

The double-population synchrotron model offers an attractive alternative to explain the X-ray emission of low redshift jets. In particular, we have tried to do a first concrete step toward the application of the shear acceleration model, for which most of the studies focused on its theoretical basis (e.g. Rieger \& Duffy 2006, 2019, Liu et al. 2017, Webb et al. 2018, 2019). Based on the observational evidence of a link between the low energy (radio, optical) and the X-ray emission, we conjecture that the pre-accelerated electrons required for the shear acceleration are energized by a shock responsible for the electrons emitting at low frequencies. Even if the shear envelops the jet for its entire length, only immediately downstream of the shocks there is a population of particles with the required energy suitable to experience fast acceleration up to multi-TeV energies and thus radiate X-rays.

We applied the model to the multifrequency emission of two knots belonging to well studied sources, PKS 0637-751 and PKS 1136-135. While the physical parameters cannot uniquely fixed, we remarked that an important constrain to the magnetic field can be derived by the total jet power required by the low-energy synchrotron emission. In particular, to keep the jet power below $P_{\rm jet}=10^{48}$ erg s$^{-1}$ fields larger than about 10 $\mu$G are required. On the other hand, substantially larger magnetic fields would determine a synchrotron cooling effectively hampering the acceleration of the electrons to the required energies. 

For both sources we found a good solution for the case $q=5/3$. For PKS 0637-752, to keep the power at acceptable values we assume that the emission is weakly beamed, quite compatible with properties of the jet. In both cases the inferred power is of the order of $10^{47}$ erg s$^{-1}$, on the high-power tail of the powerful quasars population (Ghisellini et al. 2014).

For definiteness we fix some of the parameters describing the system to fiducial values. In particular, the steepness of the velocity gradient characterizing the shear layer ($\sim \beta_{j,0}/\Delta L$) could be increase, with a corresponding reduction of the acceleration time. In principle this could be achieved either increasing the velocity jump (i.e increasing $\beta_{j,0}$) or shrinking the layer (reducing $\Delta L$). In the latter case, however, the maximum energy would be limited, instead than by acceleration efficiency, by the condition $\lambda < \Delta L$.  

To produce an appreciable flux for the high-energy bump, as required by the intense X-ray emission, we need to assume that acceleration lasts for times of the order of few times $r_j/c$, i.e. $10^3$ years, quite compatible with the expected lifetime of the sources. There is some degeneracy between $\tau_{\rm inj} $ and $t$, although for PKS 0637-752 the optical emission favors large $\tau_{\rm inj}$, since a too large injection would result in an optical continuum exceeding the observed optical flux. Therefore, in this scheme, the structures responsible for the particle acceleration (shocks and shear layers) should be relatively long-lived and we do not expect to observe variability on small timescales.

In all cases the cut-off Lorentz factor for the low-energy component associated to the shock is around $10^6$, i.e. at TeV energies. This value depends relatively weakly on the assumed magnetic field. The shape of the spectrum strongly suggest that the cut-off marks the maximum energy achieved by the acceleration mechanism, possibly related to the balance between gain and synchrotron radiative losses. In the tiny magnetic field derived here the synchrotron cooling length of these electrons is of the order of 1 Mpc. Clearly some other mechanisms is required to quench the emission. Adiabatic losses are expected to operate on timescales of the order of $r_j/c$, therefore automatically limiting the knot extension to length comparable to the jet radius (e.g. Tavecchio et al. 2003).
 
Of course, our scheme adopted several simplifications and some {\it caveats} are in order. The model is based on the non-relativistic analytical treatment of the acceleration process using a quite idealized geometry (i.e. a core surrounded by a thin layer with a linear decrease of the speed). Turbulence with standard spectra is considered to provide the requested scattering of the accelerating particles but it is not self-consistently derived. A related topic is the treatment of particle diffusion (and the already mentioned estimate of the escape time). It is likely that accelerated particles do not homogeneously fill the jet, but some kind of energy-dependent stratification is naturally produced. This, besides determining an energy-dependent volume for the emitting particles, could have an important impact if the Lorentz factor of the central core region, $\Gamma_{j,0}$, is large and therefore beaming plays a role. Due to the non-relativistic nature of the model, we limit $\Gamma_{j,0}$ to 1.7, but larger values are conceivable (and likely required if diffusive escape is efficient, Rieger \& Duffy 2019). In order to extend the model including some of the points noted above, besides the analytical studies (Rieger \& Duffy 2006, Webb et al. 2018, 2019), improved simulations (e.g.  Sironi et al. 2020, Borse et al. 2020 for recent studies) should be applied to investigate particle transport, instabilities and turbulence development in a shearing jet.

The assumption that the magnetic field assumes the same value throughout the entire flow is clearly central in the application of the constraint from the jet power. On the other hand, it is likely that the development of Kelvin-Helmholtz instability leads to the amplification of the field in the shear layer through dynamo processes (e.g. Zhang et al. 2009). Our treatment -- now imposing that the magnetic field in the downstream region is not larger than in the shear layer -- can be straightforwardly generalized to this case. Other possible complications are related to the self-generation of magnetic fields by streaming particles close to the shock front (e.g. Tavecchio et al. 2018) or the presence of strong turbulence in the downstream flow (e.g. Baring et al. 2017), possibly leading to particle reacceleration.

It has to be remarked that in comparing the detailed morphology of knots and their multifrequency appearance with the result of our simple model all the above mentioned caveats and complexities should be considered. For instance, some jets exhibit knots for which the radio emission is systematically shifted downstream with respect to the maximum of the X-ray emission (e.g. Clautice et al. 2016). While the simplest version of the shear model developed here would suggest a strict spatial coincidence between radio-optical and X-ray emission (or perhaps a small shift of the X-ray emission), the possible stochastic reacceleration/heating of low-energy electrons in the shock-induced turbulence could result into further, possibly dominant, contributions to the radio emission shifted toward the downstrem regions. Dedicated kinetic shock simulations able to capture relatively large regions would be required to self-consistently investigate the development of the turbulence and the consequent reacceleration of particles far downstream from the shock. Another effect potentially leading to the displacement of the radio emission is the expansion and the bulk reacceleration of the flow following a recollimation shock, that under a favorable geometry could result in a larger beaming amplification of the downstream emission (see e.g. Bodo \& Tavecchio 2018). 

Another potential difficulty for the model  is represented by some cases in which the transverse profile of the X-ray emission appears narrower than that of the radio (e.g. Kataoka et al. 2008). In fact, since in the model the X-rays are produced in the sheath, one would naturally expect a broader extension at the highest frequencies.  As above, a possible solution could be the re-expansion of the flow after a recollimation shock. Indeed, in the case reported by Kataoka et al. (2008) the radio emission peaks after the X-ray one, as expected in this case. The observed radio emission, therefore, would be dominated by the portion of the jet after the shock, where expansion of the flow could account for the broader intensity profile. It is however clear that, although the model offers an attractive theoretical framework, in its detailed application to specific cases one should consider several interlaced aspects that deserve further investigation and are clearly beyond the aim of this work.

Finally we would like to note that the range of parameters we have derived here is not promising for the possible acceleration of ultra-high cosmic rays. As discussed by Rieger \& Duffy (2019), for strong turbulence CR accelerated in the shear layer can reach energies of the order of $E=3\times 10^{18} Z (B/30\, \mu{\rm G}) (\Delta L/0.1 \, {\rm kpc})$ eV. Assuming $B$ of the order of 10 $\mu$G, even iron nuclei are not expected to reach energies larger than 30 EeV. This constraint (basically deriving from the low magnetic field we estimate for the acceleration region, in turn determined by the relatively large acceleration time) can be possibly by-passed if the flow is relativistic. However, a detailed modeling of the observed emission, along the lines of what we attempted here for the non-relativistic case, is required to check this possibility.

\section*{Acknowledgments}
I thank the anonymous referee for useful criticism. I am grateful to Frank Rieger for discussions and to Lukas Merten, Gabriele Ghisellini, Matteo Lucchini and Dan Schwartz for comments. FT acknowledges contribution from the grant INAF Main Stream project "High-energy extragalactic astrophysics: toward the Cherenkov Telescope Array"

\section*{Data availability}

Data available on request.


\begin{thebibliography}{}

\bibitem[\protect\citeauthoryear{Atoyan \& Dermer}{2004}]{2004ApJ...613..151A} Atoyan A., Dermer C.~D., 2004, ApJ, 613, 151

\bibitem[\protect\citeauthoryear{Baring, B{\"o}ttcher, \& Summerlin}{2017}]{2017MNRAS.464.4875B} Baring M.~G., B{\"o}ttcher M., Summerlin E.~J., 2017, MNRAS, 464, 4875

\bibitem[\protect\citeauthoryear{Blandford \& Eichler}{1987}]{1987PhR...154....1B} Blandford R., Eichler D., 1987, Physical Reports, 154, 1

\bibitem[\protect\citeauthoryear{Blandford, Meier \& Readhead}{2019}]{2019ARA&A..57..467B} Blandford R., Meier D., Readhead A., 2019, ARA\&A, 57, 467

\bibitem[\protect\citeauthoryear{Bodo \& Tavecchio}{2018}]{2018A&A...609A.122B} Bodo G., Tavecchio F., 2018, A\&A, 609, A122

\bibitem[\protect\citeauthoryear{Borse et al.}{2020}]{2020arXiv200913540B} Borse N., Acharya S., Vaidya B., Mukherjee D., Bodo G., Rossi P., Mignone A., 2020, MNRAS, submitted (arXiv:2009.13540)

\bibitem[\protect\citeauthoryear{Breiding et al.}{2017}]{2017ApJ...849...95B} Breiding P., Meyer E.~T., Georganopoulos M., Keenan M.~E., DeNigris N.~S., Hewitt J., 2017, ApJ, 849, 95

\bibitem[\protect\citeauthoryear{Cara et al.}{2013}]{2013ApJ...773..186C} Cara M., Perlman E.~S., Uchiyama Y., Cheung C.~C., Coppi P.~S., Georganopoulos M., Worrall D.~M., et al., 2013, ApJ, 773, 186

\bibitem[\protect\citeauthoryear{Celotti, Ghisellini, \& Chiaberge}{2001}]{2001MNRAS.321L...1C} Celotti A., Ghisellini G., Chiaberge M., 2001, MNRAS, 321, L1

\bibitem[\protect\citeauthoryear{Chang \& Cooper}{1970}]{1970JCoPh...6....1C} Chang J.~S., Cooper G., 1970, JCoPh, 6, 1

\bibitem[\protect\citeauthoryear{Chartas et al.}{2000}]{2000ApJ...542..655C} Chartas G., Worrall D.~M., Birkinshaw M., Cresitello-Dittmar M., Cui W., Ghosh K.~K., Harris D.~E., et al., 2000, ApJ, 542, 655

\bibitem[\protect\citeauthoryear{Clautice et al.}{2016}]{2016ApJ...826..109C} Clautice D., Perlman E.~S., Georganopoulos M., Lister M.~L., Tombesi F., Cara M., Marshall H.~L., et al., 2016, ApJ, 826, 109

\bibitem[\protect\citeauthoryear{Coppi \& Blandford}{1990}]{1990MNRAS.245..453C} Coppi P.~S., Blandford R.~D., 1990, MNRAS, 245, 453

\bibitem[\protect\citeauthoryear{Fabian}{2012}]{2012ARA&A..50..455F} Fabian A.~C., 2012, ARA\&A,

\bibitem[\protect\citeauthoryear{Georganopoulos, Meyer, \& Perlman}{2016}]{2016Galax...4...65G} Georganopoulos M., Meyer E., Perlman E., 2016, Galax, 4, 65

\bibitem[\protect\citeauthoryear{Ghisellini \& Celotti}{2001}]{2001MNRAS.327..739G} Ghisellini G., Celotti A., 2001, MNRAS, 327, 739

\bibitem[\protect\citeauthoryear{Ghisellini, Tavecchio, \& Chiaberge}{2005}]{ii} Ghisellini G., Tavecchio F., Chiaberge M., 2005, A\&A, 432, 401

\bibitem[\protect\citeauthoryear{Ghisellini et al.}{2014}]{2014Natur.515..376G} Ghisellini G., Tavecchio F., Maraschi L., Celotti A., Sbarrato T., 2014, Natur, 515, 376

\bibitem[\protect\citeauthoryear{Harris \& Stern}{1987}]{1987ApJ...313..136H} Harris D.~E., Stern C.~P., 1987, ApJ, 313, 136

\bibitem[\protect\citeauthoryear{Harris \& Krawczynski}{2002}]{2002ApJ...565..244H} Harris D.~E., Krawczynski H., 2002, ApJ, 565, 244

\bibitem[\protect\citeauthoryear{Jester et al.}{2002}]{2002A&A...385L..27J} Jester S., R{\"o}ser H.-J., Meisenheimer K., Perley R., 2002, A\&A, 385, L27

\bibitem[\protect\citeauthoryear{Kataoka \& Stawarz}{2005}]{2005ApJ...622..797K} Kataoka J., Stawarz {\L}., 2005, ApJ, 622, 797

\bibitem[\protect\citeauthoryear{Kataoka et al.}{2008}]{2008ApJ...685..839K} Kataoka J., Stawarz {\L}., Harris D.~E., Siemiginowska A., Ostrowski M., Swain M.~R., Hardcastle M.~J., et al., 2008, ApJ, 685, 839

\bibitem[\protect\citeauthoryear{Komissarov}{1994}]{1994MNRAS.266..649K} Komissarov S.~S., 1994, MNRAS, 266, 649

\bibitem[\protect\citeauthoryear{Liu, Rieger, \& Aharonian}{2017}]{2017ApJ...842...39L} Liu R.-Y., Rieger F.~M., Aharonian F.~A., 2017, ApJ, 842, 39

\bibitem[\protect\citeauthoryear{Lovell et al.}{2000}]{2000aprs.conf..215L} Lovell J.~E.~J., Tingay S.~J., Piner B.~G., Jauncey D.~L., Preston R.~A., Murphy D.~W., McCulloch P.~M., et al., 2000, in Astrophysical Phenomena Revealed by Space VLBI, ed. H. Hirabayashi, P. G. Edwards, \& D. W. Murphy (Sagamihara: ISAS), 215

\bibitem[\protect\citeauthoryear{Lucchini, Tavecchio, \& Ghisellini}{2017}]{2017MNRAS.466.4299L} Lucchini M., Tavecchio F., Ghisellini G., 2017, MNRAS, 466, 4299

\bibitem[\protect\citeauthoryear{Maraschi \& Tavecchio}{2003}]{2003ApJ...593..667M} Maraschi L., Tavecchio F., 2003, ApJ, 593, 667

\bibitem[\protect\citeauthoryear{Marshall et al.}{2005}]{2005ApJS..156...13M} Marshall H.~L., Schwartz D.~A., Lovell J.~E.~J., Murphy D.~W., Worrall D.~M., Birkinshaw M., Gelbord J.~M., et al., 2005, ApJS, 156, 13

\bibitem[\protect\citeauthoryear{Marshall et al.}{2018}]{2018ApJ...856...66M} Marshall H.~L., Gelbord J.~M., Worrall D.~M., Birkinshaw M., Schwartz D.~A., Jauncey D.~L., Griffiths G., et al., 2018, ApJ, 856, 66

\bibitem[\protect\citeauthoryear{Marshall et al.}{2011}]{2011ApJS..193...15M} Marshall H.~L., Gelbord J.~M., Schwartz D.~A., Murphy D.~W., Lovell J.~E.~J., Worrall D.~M., Birkinshaw M., et al., 2011, ApJS, 193, 15

\bibitem[\protect\citeauthoryear{Mehta et al.}{2009}]{2009ApJ...690.1706M} Mehta K.~T., Georganopoulos M., Perlman E.~S., Padgett C.~A., Chartas G., 2009, ApJ, 690, 1706

\bibitem[\protect\citeauthoryear{Meyer et al.}{2017}]{2017ApJ...835L..35M} Meyer E.~T., Breiding P., Georganopoulos M., Oteo I., Zwaan M.~A., Laing R., Godfrey L., et al., 2017, ApJL, 835, L35

\bibitem[\protect\citeauthoryear{Meyer et al.}{2015}]{2015ApJ...805..154M} Meyer E.~T., Georganopoulos M., Sparks W.~B., Godfrey L., Lovell J.~E.~J., Perlman E., 2015, ApJ, 805, 154

\bibitem[\protect\citeauthoryear{Meyer \& Georganopoulos}{2014}]{2014ApJ...780L..27M} Meyer E.~T., Georganopoulos M., 2014, ApJL, 780, L27. doi:10.1088/2041-8205/780/2/L27

\bibitem[\protect\citeauthoryear{Meyer et al.}{2019}]{2019ApJ...883L...2M} Meyer E.~T., Iyer A.~R., Reddy K., Georganopoulos M., Breiding P., Keenan M., 2019, ApJL, 883, L2

\bibitem[\protect\citeauthoryear{Merten et al.}{2020}]{pippo} Merten L.,  Boughelilba, M., Reimer A., Da Vela P., Vorobiov S., Tavecchio F., Bonnoli G. et al., 2020, Astropart. Phys. submitted

\bibitem[\protect\citeauthoryear{Mondal \& Gupta}{2019}]{2019APh...107...15M} Mondal S., Gupta N., 2019, APh, 107, 15

\bibitem[\protect\citeauthoryear{Perlman et al.}{2020}]{2020Galax...8...71P} Perlman E.~S., Clautice D., Avachat S., Cara M., Sparks W.~B., Georganopoulos M., Meyer E., 2020, Galax, 8, 71

\bibitem[\protect\citeauthoryear{Rieger \& Duffy}{2006}]{2006ApJ...652.1044R} Rieger F.~M., Duffy P., 2006, ApJ, 652, 1044

\bibitem[\protect\citeauthoryear{Rieger \& Duffy}{2019}]{2019ApJ...886L..26R} Rieger F.~M., Duffy P., 2019, ApJL, 886, L26

\bibitem[\protect\citeauthoryear{Rieger}{2019}]{2019Galax...7...78R} Rieger F.~M., 2019, Galax, 7, 78

\bibitem[\protect\citeauthoryear{Romero et al.}{2017}]{2017SSRv..207....5R} Romero G.~E., Boettcher M., Markoff S., Tavecchio F., 2017, SSRv, 207, 5

\bibitem[\protect\citeauthoryear{Roser et al.}{2000}]{2000A&A...360...99R} R{\"o}ser H.-J., Meisenheimer K., Neumann M., Conway R.~G., Perley R.~A., 2000, A\&A, 360, 99

\bibitem[\protect\citeauthoryear{Simionescu et al.}{2016}]{2016ApJ...816L..15S} Simionescu A., Stawarz {\L}., Ichinohe Y., Cheung C.~C., Jamrozy M., Siemiginowska A., Hagino K., et al., 2016, ApJL, 816, L15

\bibitem[\protect\citeauthoryear{Sambruna et al.}{2002}]{2002ApJ...571..206S} Sambruna R.~M., Maraschi L., Tavecchio F., Urry C.~M., Cheung C.~C., Chartas G., Scarpa R., et al., 2002, ApJ, 571, 206

\bibitem[\protect\citeauthoryear{Sambruna et al.}{2004}]{2004ApJ...608..698S} Sambruna R.~M., Gambill J.~K., Maraschi L., Tavecchio F., Cerutti R., Cheung C.~C., Urry C.~M., et al., 2004, ApJ, 608, 698

\bibitem[\protect\citeauthoryear{Sambruna et al.}{2006}]{2006ApJ...641..717S} Sambruna R.~M., Gliozzi M., Donato D., Maraschi L., Tavecchio F., Cheung C.~C., Urry C.~M., et al., 2006, ApJ, 641, 717

\bibitem[\protect\citeauthoryear{Sambruna \& Harris}{2012}]{2012rjag.book..185S} Sambruna R., Harris D.~E., 2012, in {\it Relativistic jets from Active Galactic Nuclei}, Eds. Boettcher, M., Harris, D.E. \& Krawczynski, H., Wiley

\bibitem[\protect\citeauthoryear{Schwartz et al.}{2000}]{2000ApJ...540L..69S} Schwartz D.~A., Marshall H.~L., Lovell J.~E.~J., Piner B.~G., Tingay S.~J., Birkinshaw M., Chartas G., et al., 2000, ApJL, 540, 69

\bibitem[\protect\citeauthoryear{Schwartz}{2002}]{2002ApJ...569L..23S} Schwartz D.~A., 2002, ApJL, 569, L23

\bibitem[\protect\citeauthoryear{Schwartz et al.}{2020}]{2020arXiv201006535S} Schwartz D., Siemiginowska A., Snios B., Worrall D., Birkinshaw M., Cheung C.~C., Marshall H., et al., 2020, ApJ, submitted (arXiv:2010.06535)

\bibitem[\protect\citeauthoryear{Siemiginowska et al.}{2007}]{2007ApJ...657..145S} Siemiginowska A., Stawarz {\L}., Cheung C.~C., Harris D.~E., Sikora M., Aldcroft T.~L., Bechtold J., 2007, ApJ, 657, 145.

\bibitem[\protect\citeauthoryear{Sironi, Rowan, \& Narayan}{2020}]{2020arXiv200911877S} Sironi L., Rowan M.~E., Narayan R., 2020, submitted (arXiv:2009.11877)

\bibitem[\protect\citeauthoryear{Stawarz \& Ostrowski}{2002}]{2002ApJ...578..763S} Stawarz {\L}., Ostrowski M., 2002, ApJ, 578, 763

\bibitem[\protect\citeauthoryear{Tavecchio, Maraschi, \& Ghisellini}{1998}]{1998ApJ...509..608T} Tavecchio F., Maraschi L., Ghisellini G., 1998, ApJ, 509, 608

\bibitem[\protect\citeauthoryear{Tavecchio et al.}{2000}]{2000ApJ...544L..23T} Tavecchio F., Maraschi L., Sambruna R.~M., Urry C.~M., 2000, ApJL, 544, L23

\bibitem[\protect\citeauthoryear{Tavecchio, Ghisellini, \& Celotti}{2003}]{2003A&A...403...83T} Tavecchio F., Ghisellini G., Celotti A., 2003, A\&A, 403, 83

\bibitem[\protect\citeauthoryear{Tavecchio}{2006}]{2006tmgm.meet..512T} Tavecchio F., 2006, in the proc. of the Xth Marcel Grossmann Meeting on General Relativity, Rio de Janeiro, Brazil, July 2003, Eds.: Mario Novello; Santiago Perez Bergliaffa; Remo Ruffini. Singapore: World Scientific Publishing

\bibitem[\protect\citeauthoryear{Tavecchio et al.}{2007}]{2007ApJ...662..900T} Tavecchio F., Maraschi L., Wolter A., Cheung C.~C., Sambruna R.~M., Urry C.~M., 2007, ApJ, 662, 900

\bibitem[\protect\citeauthoryear{Tavecchio et al.}{2018}]{2018MNRAS.480.2872T} Tavecchio F., Landoni M., Sironi L., Coppi P., 2018, MNRAS, 480, 2872

\bibitem[\protect\citeauthoryear{Uchiyama et al.}{2005}]{2005ApJ...631L.113U} Uchiyama Y., Urry C.~M., Van Duyne J., Cheung C.~C., Sambruna R.~M., Takahashi T., Tavecchio F., et al., 2005, ApJL, 631, L113

\bibitem[\protect\citeauthoryear{Uchiyama et al.}{2006}]{2006ApJ...648..910U} Uchiyama Y., Urry C.~M., Cheung C.~C., Jester S., Van Duyne J., Coppi P., Sambruna R.~M., et al., 2006, ApJ, 648, 910

\bibitem[\protect\citeauthoryear{Uchiyama et al.}{2007}]{2007ApJ...661..719U} Uchiyama Y., Urry C.~M., Coppi P., Van Duyne J., Cheung C.~C., Sambruna R.~M., Takahashi T., et al., 2007, ApJ, 661

\bibitem[\protect\citeauthoryear{Webb et al.}{2018}]{2018ApJ...855...31W} Webb G.~M., Barghouty A.~F., Hu Q., le Roux J.~A., 2018, ApJ, 855, 31

\bibitem[\protect\citeauthoryear{Webb et al.}{2019}]{2019ApJ...881..123W} Webb G.~M., Al-Nussirat S., Mostafavi P., Barghouty A.~F., Li G., le Roux J.~A., Zank G.~P., 2019, ApJ, 881, 123

\bibitem[\protect\citeauthoryear{Zhang, MacFadyen, \& Wang}{2009}]{2009ApJ...692L..40Z} Zhang W., MacFadyen A., Wang P., 2009, ApJL, 692, L40

\end{thebibliography}
\end{document}